# The Orbital Period of the Ultraluminous X-Ray Source in M82

**Philip Kaaret\*, Melanie G. Simet, and Cornelia C. Lang**

Department of Physics and Astronomy, University of Iowa, Van Allen Hall, Iowa City, IA 52242, USA

\*To whom correspondence should be addressed.

E-mail: philip-kaaret@uiowa.edu

Intermediate-mass black holes (*1*), which have masses in the range $10^2$ to $10^4$ solar masses ($M_\odot$) -- larger than can be produced in the collapse of a single normal star and smaller than the supermassive black holes found in galactic nuclei -- would be a new class of astrophysical object and may be important in the formation of supermassive black holes (*2*). The brightest x-ray source in the nearby starburst galaxy M82 has an apparent luminosity reaching $10^{41}$ erg s$^{-1}$, indicating a mass exceeding 500 $M_\odot$ (*2,3*). Its displacement from the galaxy nucleus indicates a mass less than $10^5 M_\odot$ because of dynamical friction (*3*). The source shows quasiperiodic x-ray oscillations at frequencies of 0.05-0.1 Hz, suggestive of a mass near 100 to 1000 $M_\odot$ (*4*).

We monitored the x-ray emission from M82 every other day for 240 days using the Proportional Counter Array on the Rossi X-Ray Timing Explorer (*5*). We found that the x-ray flux from M82 is modulated (Fig. 1) with a peak to peak amplitude corresponding to an isotropic luminosity of $2.4\times10^{40}$ erg s$^{-1}$ in M82 and a period of $62.0 \pm 2.5$ days. The peak arrival times appear periodic to the accuracy of measurement.

X-ray modulations at the orbital period are known from several black hole X-ray binaries (*6*). Therefore, we interpret the x-ray periodicity from M82 as the orbital period of the ultraluminous x-ray source (ULX). Superorbital modulations in black hole X-ray binaries occur with periods of 162 to 600 days, which are longer than the observed period.

The high inferred luminosity indicates that the black hole is gravitationally pulling mass directly from the outer surface of its companion star via Roche-lobe overflow (*7*). Because of the geometry of the binary system, the orbital period is related to the companion star density $\rho \approx 115\, P^{-2}$ g cm$^{-3}$ where *P* is the period in hours (*8*). For a 62 day orbital period, the mean density of the companion star is $5\times10^{-5}$ g cm$^{-3}$. This excludes normal main sequence stars, which are denser than $10^{-2}$ g cm$^{-3}$, but is compatible with giant or supergiant stars.

The ULX in M82 lies close to (as projected on the sky) and possibly within the super star cluster MGG 11 (*7,9*). An intermediate mass black hole may have been formed by stellar collisions in the extremely dense core of MGG 11 (*9*). Infrared spectroscopy of MGG 11 shows that its near-infrared light is dominated by red supergiant stars (*10,11*). Hence, the cluster has existed long enough for the ULX companion star to evolve to the giant stage.

When the companion star of an intermediate mass black hole evolves through the giant phase, mass transfer causes the orbit to widen and the orbital period increases through 10-100 days as the mass transfer rate reaches $10^{-4}$ $M_\odot$ yr$^{-1}$ for an initial companion mass near 15 $M_\odot$ (*12,13*). This mass transfer rate is sufficient to power an X-ray luminosity of $10^{41}$ erg s$^{-1}$. The lifetime of this phase is short, of order $10^5$ yr, compared to the companion age of $10^7$ yr, suggesting that we are fortunate to see the M82 ULX during a brief and unusually bright phase of its evolution and that other ULXs reaching such extreme luminosities, near $10^{41}$ erg s$^{-1}$, should be rare.

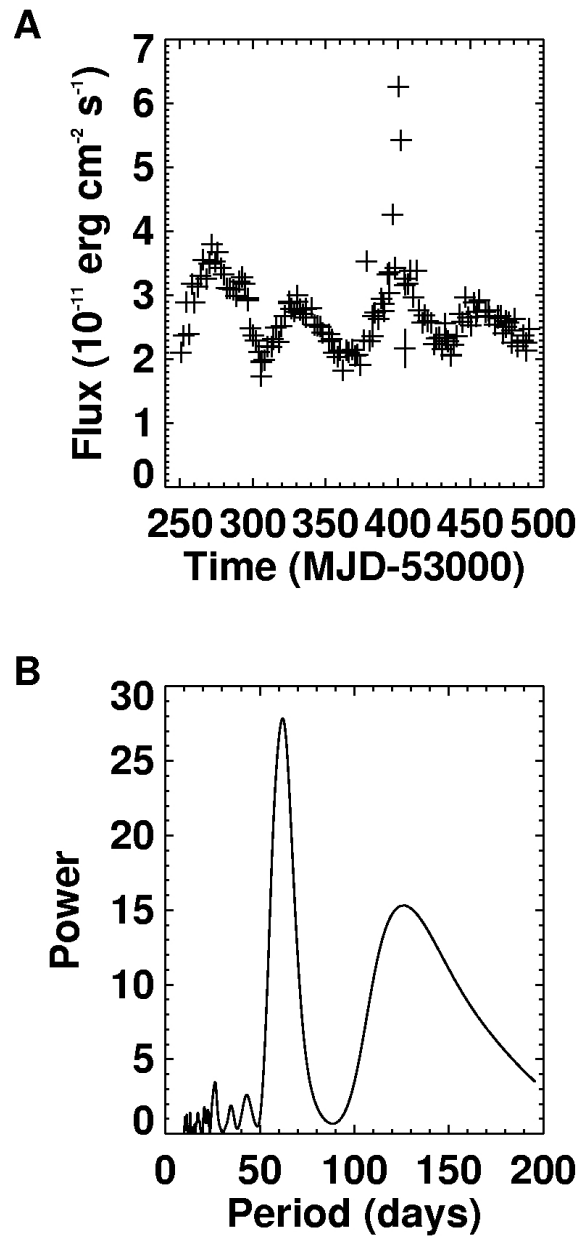

**Fig. 1** (**A**) X-ray light curve of M82. The X-ray flux in the 2-10 keV band from M82 measured using the Proportional Counter Array on the Rossi X-Ray Timing Explorer is shown versus time in modified Julian days (MJD). A periodicity near 60 days is evident with a peak-to-peak flux of $1.5 \times 10^{-11}$ erg cm$^{-1}$ s$^{-1}$. There is an x-ray flare at MJD 53400.2. (**B**) A periodogram of the data in (**A**) and two earlier data points from MJD 53098.9 and 51153.7. The peak at $62.0 \pm 2.5$ days has a power of 27.9 (*14*). The chance probability of occurrence, taking into account the number of trials, is $1.1 \times 10^{-10}$. Removing the two early data points or points with fluxes above $4 \times 10^{-11}$ erg cm$^{-1}$ s$^{-1}$ does not significantly shift the peak. There is a secondary peak at 125.7 days with a power of 15.3.